\documentclass[twocolumn,showpacs,tighten,floats,amsmath,amssymb,prb]{revtex4}
\usepackage{graphicx}
\usepackage{amssymb}
\usepackage{epsfig}      
\usepackage{dcolumn}     
\usepackage{bm}          
\def\rcoo{$r_{Cu-O1}$}
\def\rcot{$r_{Cu-O2}$}
\def\rlot{$r_{La/Sr-O2}$}
\def\uot{$U_{33}$(O2)}

\def\sb#1{$_{#1}$}

\def\aporb{Cu{$d_{3z^2-r^2}$}--O{$p_{z}$}}
\def\plorb{Cu{$d_{x^2-y^2}$}--O{$p_{x,y}$}}
\def\dxy{$d_{x^2-y^2}$}
\def\lsco{La$_{2-x}$Sr$_x$CuO$_4$}

\def\etal{{\it et al.}}
\begin{document}
\title{Nominal doping and partition of doped holes between planar and
apical orbitals in {\lsco }}
\author{E.~S. {Bo\v zin} and S.~J.~L. Billinge}
\affiliation{Department of Physics and Astronomy, Michigan State University, East Lansing, MI 48824}
\email{bozin@pa.msu.edu}
\homepage{http://nirt.pa.msu.edu/}
\date{\today}
\begin{abstract}
By considering both the average structural parameters obtained from Rietveld 
refinement of neutron powder diffraction data, and the local structural 
parameters obtained from the atomic pair distribution function, we have tested 
the recent hypothesis of Perry \etal\ [Perry et al., Phys. Rev. B {\bf 65}, 
144501 (2002)] that doping in \lsco\ system occurs as localized defects of 
predominantly \aporb\ character associated with the Sr dopants accompanied 
by a local destruction of the Jahn-Teller distortion. While the structural 
parameters behave qualitatively according to the prediction of this model, 
a quantitative analysis indicates that doped holes predominantly appear in 
the planar \plorb\ band as is normally assumed.  However, a small amount of 
the doped charge does enter the \aporb\ orbitals and this should be taken 
into account when theoretical phase diagrams are compared to experiment. 
We present a calibration curve, $p = x (1.00(1) - 0.45(7) x)$, for the planar 
charge doping, $p$, vs strontium content, $x$, for the \lsco\ system.
\end{abstract}
\maketitle
\section{
Introduction
}

Cuprate high temperature superconductors are doped Mott insulators. The novel 
superconductivity appears at doping levels just beyond the insulator-metal 
(IM) transition. The insulating behavior of the undoped endmember is 
understood to be due to electron correlation effects in the half-filled 
planar Cu\dxy --O$p_\alpha$ ($\alpha=x, y$) anti-bonding band. 
The phase diagram of the cuprates can be interpreted in terms of holes doped 
into this planar band. For example, in the system \lsco\ it is assumed that 
one hole enters this band per strontium atom.  This paradigm for the doping 
has hardly been questioned and a multitude of papers exist implicitly 
assuming this behavior. The \lsco\ system is archetypal since it is a single 
layer system which can be straightforwardly doped over a wide range. The 
resulting phase diagram is thought to exhibit features that are universal 
to the cuprates. 
However, as a result of recent 
{\it ab initio} electronic band structure calculations, Perry, Tahir-Kheli and 
Goddard (PTG)~\cite{perry;prb02} have suggested 
a new model for the doping in the \lsco\ system. In this case the doped holes 
reside in localized states on the CuO$_6$ octahedra situated next to the 
dopant strontium ions. Furthermore, the doped charge resides principally in 
the \aporb\ orbitals (i.e., the out-of-plane bonds). Coincidentally 
a structural distortion occurs such that the local Jahn-Teller (JT) 
distortion, that results in the long Cu-O apical bond, is destroyed. A flat 
impurity band forms in the gap. The IM transition then occurs on increased 
doping in a manner similar to doped semiconductors by percolation of 
the doped impurities. These calculations were originally motivated by 
XAFS measurements that observed a distorted environment of the octahedra 
in the vicinity of dopant ions~\cite{haske;prb97}. 
This new way of understanding the doping, if it is right, clearly will 
result in a paradigm shift in our understanding of cuprate physics. 
It is thus of the greatest importance to test the hypothesis 
experimentally. A rather direct probe of this doping mechanism is the 
structure because it involves the local destruction of JT elongated bonds. 
The $average$ Cu-O bond length along $z$ will thus shorten with doping and 
the width of the bond-length distribution for this bond will increase 
reflecting the increased disorder. Despite detailed structural studies of 
this system~\cite{radae;prb94i} the temperature and doping dependence of 
these parameters has not been reported. In this paper we investigate
whether there is evidence supporting this new doping paradigm 
in both the average crystal structure, and the local structure as measured 
using the  atomic pair distribution function (PDF) 
technique~\cite{billi;cc04}. We reexamine the extensive earlier structural 
data of Radaelli to extract the desired parameters.  We have also analyzed 
new neutron powder diffraction data using Rietveld refinement and PDF 
refinement. We find that the data qualitatively agree with the predictions 
of the Perry~\etal\cite{perry;prb02}. However, a more quantitative analysis, 
including evaluating the bond valences as a function of doping of the
in-plane and apical Cu-O bonds, indicates that doped charge is
predominantly residing in the planar bonds suggesting the existing
paradigm for doping is valid.

\section{
Experimental
}

\begin{figure}[tb]
\begin{center}$\,$
\includegraphics[width=3.0in,keepaspectratio=1]{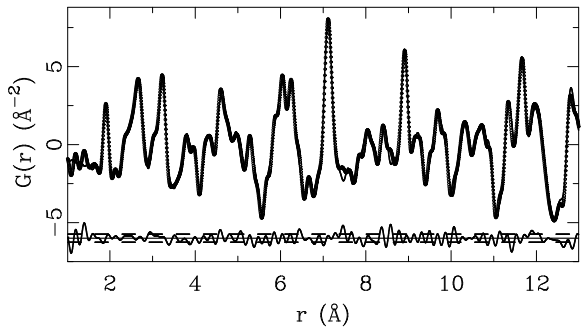}
\end{center}
%[Representative full profile PDF fit.]
\caption{Representative full profile
PDF fit of standard LTO model to 10~K data for undoped sample. Experimental
profile shown as open circles, model profile as solid line. Difference curve
is shown below the fit as solid line. Dashed lines denote experimental
uncertainties at the $2\sigma$ level. PDFs studied here utilize
diffraction information up to $Q_{MAX}=35$~\AA\ $^{-1}$.
}
\protect\label{fig;fig0}
\end{figure}

This study reexamines the extensive published data of 
Radaelli~\etal\cite{radae;prb94i} 
to extract as a function of doping the interesting parameters of the 
plane copper to apical oxygen bond length, \rcot ~and the short 
lanthanum/strontium to apical oxygen bond length, \rlot . 
These parameters were determined from the reported 
fractional coordinates and lattice parameters. 
Also important is the doping dependence of the copper to in-plane 
oxygen bond length, \rcoo , which was presented 
previously~\cite{radae;prb94i} but is reexamined 
here. To check the PTG prediction we also need the anisotropic displacement 
parameter
along the $z$-axis of the apical oxygen, \uot .  This was not published 
in~\cite{radae;prb94i} so we have used recently collected neutron powder 
diffraction data of our own from a less extensive set of samples.  Samples 
were synthesized using solid state methods and loose powders of $\sim 10$~g 
sealed in vanadium cylinders were measured at 10~K a the GEM diffractometer 
at the ISIS neutron source.  Details of sample synthesis, characterization 
and measurement are reported  elsewhere~\cite{bozin;prb99,bozin;phd03}. 
Rietveld refinements were carried out on the data using the program 
GSAS~\cite{larso;laur94}. The data were also corrected for experimental 
effects and normalized to obtain the atomic pair distribution 
function~\cite{egami;b;utbp03} (PDF) using the program 
PDFgetN~\cite{peter;jac00}.  The PDF is the Fourier transform of the 
normalized corrected powder diffraction data.  It utilizes both Bragg and 
diffuse scattering and contains information about the local 
structure~\cite{billi;cc04,egami;b;utbp03}. 
Structural models were refined to the 
PDFs using the program PDFFIT~\cite{proff;jac99}. To check for consistency, 
the same parameters were varied in the PDF and Rietveld refinements. 
An example of the measured PDF from the undoped sample is shown
in Figure~\ref{fig;fig0} with the best fit model-PDF plotted on top. 
The PDF allows us to explore models that contain non-periodic defects such 
as those proposed by PTG.  The PDF was calculated from a supercell model 
with the doping induced defects discussed by PTG.  This was then used as 
simulated data and fit using the usual long-range orthorhombic 
(LTO)~\cite{radae;prb94i} structural model.  In this way it was possible to 
estimate the size of the enlarged displacement factors that would result from 
the presence of the doping induced disorder described by PTG.  In 
Perry \etal~\cite{perry;prb02}, for simplicity, no octahedral tilts 
were considered.  In the present case we must compare resulting models with 
real data and so the doping induced defects described in~\cite{perry;prb02} 
were superimposed on the background of octahedral tilts observed in the 
undoped endmember.
%%%%%%%%%%%%%%%%%%%%%%%%%%%%%%%%%%%%%%%%%%%%%%%%%%%%%%%%%%%%%%%%
%       FIGURE
%%%%%%%%%%%%%%%%%%%%%%%%%%%%%%%%%%%%%%%%%%%%%%%%%%%%%%%%%%%%%%%%
%
%
%%%%%%%%%%%%%%%%%%%%%%%%%%%%%%%%%%%%%%%%%%%%%%%%%%%%%%%%%%%%%%%%
%       FIGURE
%%%%%%%%%%%%%%%%%%%%%%%%%%%%%%%%%%%%%%%%%%%%%%%%%%%%%%%%%%%%%%%%

\section{
Results and Discussion
}

\begin{figure}[tb]
\begin{center}$\,$
\includegraphics[height=3.75in,keepaspectratio=1]{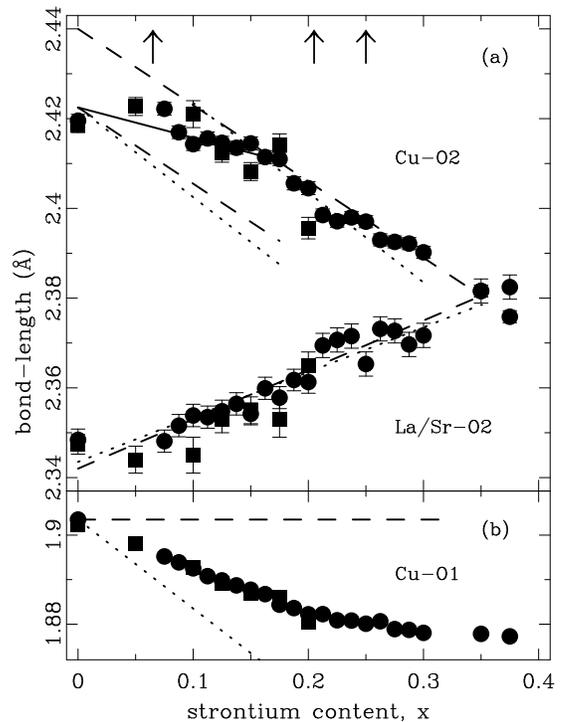}
\end{center}
\caption{Evolution of $r_{Cu-O2}$ and $r_{La/Sr-O2}$ (a), and $r_{Cu-O1}$ 
(b) average bond lengths with hole doping at 10~K: Rietveld result 
from\protect\cite{radae;prb94i} (solid circles) and PDF 
result (solid squares). Dotted lines: slope predicted from 
simple electrostatics considerations. Dashed lines: slope 
prediction based on PTG model. Arrows (from left to right): IM transition, 
structural phase transition, and disappearance of superconductivity. 
}
\protect\label{fig;fig1}
\end{figure}

\begin{figure}[tb]
\begin{center}$\,$
\includegraphics[width=2.8in,keepaspectratio=1]{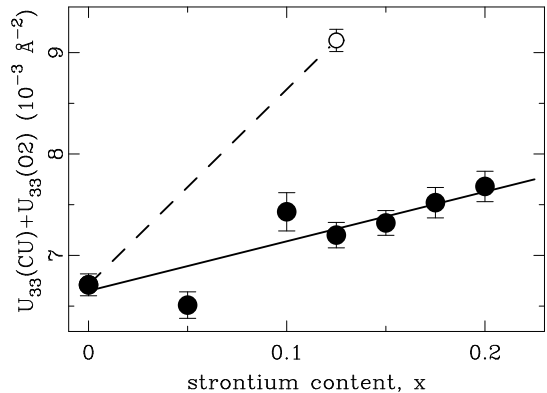}
\end{center}
\caption{Sum of displacement parameters $U_{33}$ for $Cu$ and $O2$ as a 
function of strontium content at 10~K, from PDF refinements. 
Solid line is a guide to the eye. Dashed line denotes expected 
increase in the PDF displacement parameters, estimated from PDF 
simulation based on PTG model for 1/8 doping (see text).
}
\protect\label{fig;fig2}
\end{figure}

The qualitative predictions for the evolution of the average structure with 
doping in the scenario of Perry~\etal~\cite{perry;prb02} are: 
(1) decrease in the average \rcot\ due to the local destruction of 
Jahn-Teller distortions at Cu$^{3+}$ sites; 
(2) increase in the \uot\ with doping because of the coexistence
of Jahn-Teller distorted and non-Jahn-Teller distorted octahedra; 
(3) increase in the average $r_{La/Sr-O(2)}$ distance. 
All these effects are seen qualitatively in the data as shown in 
Figs.~\ref{fig;fig1} and \ref{fig;fig2}.  

%%
%%%%%%%%%%%%%%%%%%%%%%%%%%%%%%%%%%%%%%%%%%%%%%%%%%%%%%%%%%%%%%%%%%%%%%%%%%%%%%%

%%%%%%%%%%%%%%%%%%%%%%%%%%%%%%%%%%%%%%%%%%%%%%%%%%%%%%%%%%%%%%%%
%       FIGURE
%%%%%%%%%%%%%%%%%%%%%%%%%%%%%%%%%%%%%%%%%%%%%%%%%%%%%%%%%%%%%%%%
%
%
%%%%%%%%%%%%%%%%%%%%%%%%%%%%%%%%%%%%%%%%%%%%%%%%%%%%%%%%%%%%%%%%
%       FIGURE
%%%%%%%%%%%%%%%%%%%%%%%%%%%%%%%%%%%%%%%%%%%%%%%%%%%%%%%%%%%%%%%%

However, similar effects can be expected even if the doping is taking place 
in the conventional way: homogeneously into the planar bonds.  For example, 
if the CuO$_2$ planes are becoming more positively charged with doping the 
negatively charged apical oxygen would be predicted to come closer to the 
planes due to simple coulomb attraction.  Also, the U$_{33}$ displacement 
factor of the apical oxygen atom is expected to increase with doping due to 
the random doping of larger Sr$^{2+}$ ions. Inhomogeneous doping in the form of stripes 
or checkerboards~\cite{komiy;prl05,hanag;nat04} would also result in enlarged 
thermal factors for in-plane and apical motions of $O1$ and $O2$. However, inhomogeneous 
doping would not affect average bond-length obtained from Rietveld refinement. 
It is therefore important to be more quantitative to distinguish these different 
possibilities.

The PTG calculations~\cite{perry;prb02} were carried out at special rational 
doping fractions of 1/8, 1/4 and 1/2 and suggest the appearance of a single 
distorted CuO$_6$ octahedron associated with each doped strontium.  The 
\rcot\ bond closest to the strontium is shortened by $\Delta$\rcot~$=-0.24$~\AA\
and that farthest from the Sr on the same octahedron by 
$\Delta$\rcot~$=-0.10$~\AA , the Sr-O2 distance increases, 
$\Delta$\rlot~$=+0.11$~\AA\ and the in-plane \rcoo\ 
distances do not change.  In the average structure these defects are not seen
explicitly; however, the distortions will be apparent as a properly weighted
change in the average bond-length and an increase in the respective 
displacement parameters.  If we assume that these defects appear as each 
strontium is doped and are not a special feature of the rational doping 
fractions studied, we get the following relations for the strontium doping, 
$x$, dependence of the average bond lengths:
\begin{equation}
    \Delta r_{Cu-O2}(x) = -0.17~x, ~~\Delta r_{La/Sr-O2}(x) = 0.11~x,
\end{equation}
\noindent where $x$ is the doping level. These are shown as the dashed lines 
in Fig.~\ref{fig;fig1}.  The evolution of \rlot\ is in quantitative agreement.
The case of \rcot\ is more complicated.  \rcot\ decreases much more slowly 
than the prediction initially on doping.  However, beyond a doping level of 
$\sim 0.17$ the slope of \rcot\ vs $x$ increases.  In this higher doping 
region the slope of the PTG line agrees rather well with the data.  There is 
no specific prediction from the PTG calculations for the $x$-dependence of 
\rcoo , although it is expected to be small since the length of \rcoo\ in the 
defect octahedra does not change and no doping dependent change in lattice 
parameters has been assumed in their calculations. This reflects the fact 
that, in the PTG picture the doped charge is almost exclusively located in 
the apical $d_{z^2-r^2}$ orbitals of the copper. Therefore, we have drawn a 
flat dashed line as the PTG prediction for this parameter in 
Fig.~\ref{fig;fig1}(b).  As is apparent, the data for \rcoo\ slope downward 
with a slope of $\Delta$\rcoo~$\sim-0.105$. This is comparable to, though 
slightly less than, the observed $\Delta$\rcot~$\sim-0.18$ of the apical bond.
This observation will be important later. 
%%%%%%%%%%%%%%%%%%%%%%%%%%%%%%%%%%%%%%%%%%%%%%%%%%%%%%%%%%%%%%%%
%       FIGURE
%%%%%%%%%%%%%%%%%%%%%%%%%%%%%%%%%%%%%%%%%%%%%%%%%%%%%%%%%%%%%%%%
%

%
%%%%%%%%%%%%%%%%%%%%%%%%%%%%%%%%%%%%%%%%%%%%%%%%%%%%%%%%%%%%%%%%
%       FIGURE
%%%%%%%%%%%%%%%%%%%%%%%%%%%%%%%%%%%%%%%%%%%%%%%%%%%%%%%%%%%%%%%%

The presence of the anti-Jahn-Teller distorted doped CuO$_6$ octahedra
in the PTG calculations will manifest itself in an increased atomic 
displacement factor for O2, especially along $z$. 
The most sensitive parameter will be the width of the Cu-O pair distribution
along the $z$-axis.  This cannot be measured directly from the PDF because of peak
overlap but can be found from the sum of U\sb{33}(Cu)+U\sb{33}(02).  The
mean-square width 
is plotted in Fig.~\ref{fig;fig2} 
 and does 
increase with increasing doping at constant temperature, a sign that static 
disorder associated with this bond is increasing with doping.  To investigate 
more quantitatively whether this increase is consistent with the PTG 
predictions we calculated the PDF of a model structure containing the PTG 
defects.  This simulated PDF was then fit using PDFFIT using the standard LTO 
model.  This gives us a measure of how much \uot\ would increase in a standard 
LTO-structure refinement due to the presence of the PTG defects.  The PDF was 
simulated for the $x=0.125$ composition using a 8x unit cell in the following 
way.  The reference structure used was that of the undoped material with 
atomic positions taken from~\cite{radae;prb94i} and displacement factors 
taken from our PDF result for the undoped sample.  One La in the supercell 
was then replaced with Sr and the atoms on the CuO$_6$ octahedron adjacent to 
the Sr were displaced by the $\Delta$-values from the PTG prediction given 
above. 
The displacement parameters in the model retained their values in the $x=0$ 
compound.  This neglects strain due to the defects, and any stripe or checkerboard 
doping inhomogeneities in the CuO$_2$ planes, that would tend to 
increase displacement parameters further.  The resulting PDFs were refined 
with the standard LTO unit cell.  This resulted in an enlarged value of  
U\sb{33}(Cu)+U\sb{33}(02)$=9.1\times 10^{-3}$~\AA $^{-2}$
plotted on Fig.~\ref{fig;fig2}, which overestimates the observed 
displacement factors. 

%%%
Next, using PDF data of 1/8 doped \lsco\ at 10~K, {\it explicit modeling} of
the anti-JT (A-JT) model was performed. The outcome is compared to the
results of the modeling of the standard LTO model. The modeling employed
the program PDFFIT~\cite{proff;jac99}. A large supercell model
was generated that incorporates explicit Sr doping, and allows for defining
special doping-affected octahedral units, as described above in accord
with Ref.~\cite{perry;prb02}. Three distinct variants of the explicit A-JT
models were attempted. These variants are distinguished as follows: (1) A
model with symmetric A-JT distortion on affected octahedral units 
(apical Cu-O distances restricted to have the same length). (2) A model with
asymmetric A-JT distortion (relevant apical distances are allowed to
vary independently), and  (3) A model with asymmetric A-JT distortion this
time restricted to have displacements along LTO $c$-axis only. In all three
variants atomic parameters of Sr were decoupled from those of La, except for
the displacement factors that were restricted to be the same. In the variant
(3) Sr $y$-coordinate was set to zero and Sr motion was restricted to be along
$c$-axis only. The A-JT models were refined from the same set of initial
values as that used for modeling the standard LTO structure. This equality of
the starting values was only violated for O2 and Sr fractional $y$-coordinates
in the variant (3), where these were set to zero and fixed.

%%%%%%%%%%%%%%%%%%%%%%%%%%%%%%%%%%%%%%%%%%%%%%%%%%%%%%%%%%%%%%%%
%       TABLE
%%%%%%%%%%%%%%%%%%%%%%%%%%%%%%%%%%%%%%%%%%%%%%%%%%%%%%%%%%%%%%%%
%
\begin{table}
  %[Standard LTO model versus A-JT defect models.]
  \caption{Standard LTO model versus A-JT defect models: PDFFIT modeling
    results summarizing relevant distances. All lengths are in~\AA. Rw is
    the weighted PDFFIT agreement factor.\smallskip}
  \begin{center}
    \begin{tabular}{|c||c|c|c|c|} \hline
      ~~~~~~&LTO&A-JT(1)&A-JT(2)&A-JT(3) \\ \hline\hline
      Rw&0.127&0.122&0.131&0.127 \\ \hline
      r(Cu-O2)&2.4179(24)&2.4095(25)&2.4119(22)&2.406(3) \\
      &-&2.460(19)&2.450(20)&2.49(4) \\
      &-&-&2.469(19)&2.51(3) \\ \hline
      r(La-O2)&2.346(3)&2.354(3)&2.354(3)&2.361(4) \\ \hline
      &-&2.305(20)&2.305(20)&2.27(4) \\
      r(Sr-O2)&2.346(3)&2.27(3)&2.27(3)&2.20(5) \\ \hline
    \end{tabular}
  \end{center}
  \label{tab;modresAJT}
\end{table}
%
%%%%%%%%%%%%%%%%%%%%%%%%%%%%%%%%%%%%%%%%%%%%%%%%%%%%%%%%%%%%%%%%
%       TABLE
%%%%%%%%%%%%%%%%%%%%%%%%%%%%%%%%%%%%%%%%%%%%%%%%%%%%%%%%%%%%%%%%

The results are summarized in Table~\ref{tab;modresAJT}, where the weighted
PDFFIT agreement factors (Rw)~\cite{proff;jac99} are reported, as well as
all relevant distances. Although for all the variants of the A-JT model
the agreement factor is similar to the one for the LTO model, and for some
variants even somewhat more favorable, it is not sufficient to unambiguously
favor the distorted model over the standard LTO structure. Moreover, the
refined distortions obtained from such explicit modeling yield relevant 
distance changes in the direction {\it opposite} to that originally proposed 
within the A-JT model of Perry~\etal~\cite{perry;prb02} (i.e., the Cu-O2 
distance of the affected octahedral unit lengthens, while $r_{Sr-O2}$ 
shortens). However, inspection of the distances for the A-JT model results 
in the proper weighted average values when compared to corresponding average
distances obtained by employment of the standard LTO model.

%%%

We now consider the expected structural changes that would occur based on 
electrostatic considerations in the conventional doping model.  First, we 
note that the charge of the CuO$_2$ plane is becoming less negative with 
doping which could result in the apical O$^{2-}$ moving closer to the plane.  
A very rough estimate of this behavior is possible neglecting the Jahn-Teller 
effect and assuming an ionic picture by using ionic radii 
of O$^{2-}=1.35$~\AA, Cu$^{2+}=0.73$~\AA, Cu$^{3+}=0.54$~\AA, 
La$^{3+}=1.16$~\AA, and Sr$^{2+}=1.26$~\AA~\cite{shann;ac69}.  From this we 
get $\Delta$\rcot~$=-0.19~x$ and $\Delta$\rlot~$=0.1~x$ which appear in 
Fig.~\ref{fig;fig1} as dotted lines.  Not surprisingly, these curves are 
similar to those predicted by PTG that is also a model involving 
a mixture of Cu$^{2+}$ and Cu$^{3+}$ sites. Uniform doping of charge into 
the Cu-O planes would not result in an increase in \uot\ on its own; however, 
the presence of misfitting Sr$^{2+}$ dopant ions would. Any inhomogeneous charge 
distribution in the plane would increase \uot\. The fact that the 
observed increase in \uot\ exists but is smaller than needed to explain the 
PTG defects may argue in favor of electrostatic interpretation.  There is a suggestion 
from single crystal refinements compared with empirical potential calculations 
that increases in displacement factors with doping can be explained in this 
way without invoking PTG-type defects~\cite{brade;prb01}.  Our results would 
tend to support this.

\begin{figure}[tb]
\begin{center}$\,$
\includegraphics[height=3.75in,keepaspectratio=1]{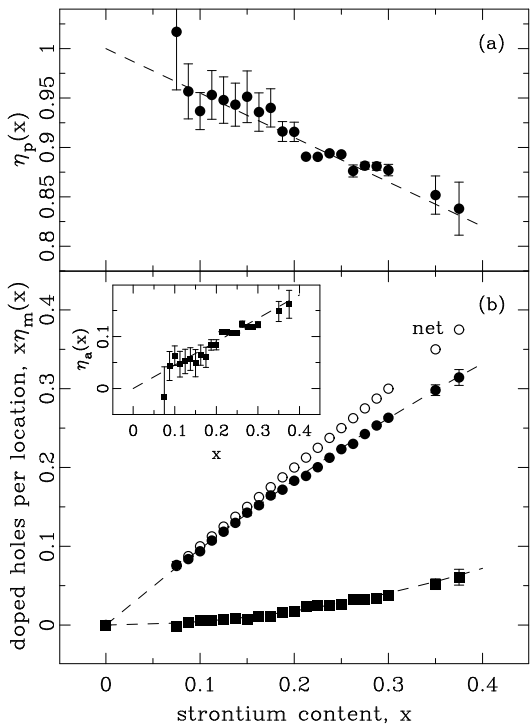}
\end{center}
\caption{(a) Partitioning parameter vs.\ Sr content. (b) Estimated average 
distribution of doped charge at 10~K based on bond valence calculations using 
Rietveld obtained distances: net apical share (solid squares), net planar 
share (solid circles), total doped charge (open circles). Inset: a measure of 
the amount of charge doped into the apical orbitals. Dashed lines represent fit to 
the data (see text for details). 
}
\protect\label{fig;fig3}
\end{figure}

A more refined empirical framework for studying the distribution of charge 
between bonds is the bond valence model~\cite{brown;acb85}.  
In this theory there is a direct relationship established between bond-length, 
$r$, and the amount of charge in a bond (the bond-valence, $s$). 
The bond valence is defined as
\begin{equation}
s(r) = \exp{(r_0-r)\over B},
\label{eq;bvs}
\end{equation}
\noindent where $B=0.37$~\AA\ is a universal constant and $r_0$ depends on
the chemical identity of the ions in question. 
Using this approach it is possible to determine how much of the doped 
charge is going into the planar vs. the apical bonds from a measurement 
of their length changes. 
The theory is successful in unstrained compounds and has
recently been applied to mixed valent crystals and crystals containing
Jahn-Teller distortions~\cite{brown;jssc89}. 
The nominal valence, $V_i$, of an ion is then determined by summing the
bond-valences over all the bonds in which it participates,
$V_i = \sum_j s_{ij}$. In the case of mixed valent crystals some ambiguity
exists as to exactly what value to use for $r_0$. The qualitative
results of our analysis, however,  do not depend on what value of $r_0$
is used. 
The results are shown in Fig.~\ref{fig;fig3}.
%%%%%%%%%%%%%%%%%%%%%%%%%%%%%%%%%%%%%%%%%%%%%%%%%%%%%%%%%%%%%%%%
%       FIGURE
%%%%%%%%%%%%%%%%%%%%%%%%%%%%%%%%%%%%%%%%%%%%%%%%%%%%%%%%%%%%%%%%
%
%
%%%%%%%%%%%%%%%%%%%%%%%%%%%%%%%%%%%%%%%%%%%%%%%%%%%%%%%%%%%%%%%%
%       FIGURE
%%%%%%%%%%%%%%%%%%%%%%%%%%%%%%%%%%%%%%%%%%%%%%%%%%%%%%%%%%%%%%%%
Here we plot how the total doped charge (assumed to be given by $x$) 
distributes itself between the apical and planar orbitals respectively, as 
determined from the bond-valence sums. The dimensionless 
variables, $\eta_p$ (solid circles) and $\eta_a$ parameterize this 
partition where $\eta_p + \eta_a=1$. 
%Details of these calculations will be presented elsewhere~\cite{bozin;phd03}

Details of these calculations are as follows. First we determine the change
in bond-valence of a particular bond due to doping,
$\delta s_i(x)=s_i(x)-s_i(0)$.  We are interested in the partition of charge
between the planar and apical orbitals, therefore we define an
orbital-valence $\delta S_m=\sum_i \delta s_i$ where the sum is over the
planar bonds for the \plorb\ orbital, $\delta S_p$, and over the apical
bonds for the \aporb\ orbital, $\delta S_a$. This is then expressed in
dimensionless parameters by dividing by the total {\it excess} valence of
copper determined by the sum over all the orbital-valences,
$\eta_m(x)=\delta S_m(x)/V_{Cu}^{ex}$.  What is plotted in
Figure~\ref{fig;fig3} is $x\eta_p$ (solid circles) and $x\eta_a$ (solid
squares), that are measures of the amount of charge doped into the
planar and apical orbitals, respectively.

As is apparent in Fig.~\ref{fig;fig1}, {\it both} the apical and in-plane 
bonds shorten with doping indicating a reduction in charge and doped holes 
going into both planar and apical bonds. However, one of the key aspects of 
this theory is the non-linear form of the bond valence, $s(r)$. 
As a result, the same amount of charge doped into a long-bond shortens it 
much more than it would a short-bond, and vice versa. At $x=0$ \rcoo$~<~$\rcot 
~reflecting the fact that initially the holes that make Cu in the 
${2+}$-state reside predominantly in the planar orbitals making these 
the canonical half-filled anti-bonding bands, with the apical bonds being 
doubly occupied with electrons (i.e., full, no holes). 
The physics of this is the existence of the Jahn-Teller distortion; however, 
this shows that the phenomenological bond-valence model is making the correct 
prediction for the charge segregation initially. Subsequently, the 
experimental observation is that {\it both the planar and 
apical bonds shorten at about the same rate} 
but because the planar bonds are initially shorter, and the fact that there 
are four of them rather than two, mean that the experimental observations 
demand that the doped charge is going predominantly into the planar 
orbitals as is evident in Fig.~\ref{fig;fig3}.  

The main result of this analysis is that the simple picture of localized 
\aporb\ doped holes associated with Sr sites described by the PTG calculations 
is not supported by the structural data. To a rather good approximation doped 
charge is going into the planar \plorb\ orbitals. We do not address here whether 
$Cu$ sites contain different amounts of charge and this does not rule out 
localized doped defects or striped or checkerboard patterns of charge if they 
have sufficient planar character. Such defects have been stabilized in the 
unrestricted Becke-3-Lee-Yang-Parr density functional calculations~\cite{takhi;unpub}. 
However, the observation that \rcot\ is getting shorter also requires that 
some doped charge is appearing in the apical \aporb\  orbital and a correction 
should be made to the canonical $p=x$ relationship to account for this, where $p$ is the doped 
charge in the planar orbitals. The partitioning parameter $\eta_p$ gives the 
correction, and $p$ is therefore given by $p=x\eta_p$ ($\eta_p$ shown in 
Fig.~\ref{fig;fig3}(a)). This augments spectroscopic measurements that show 
the partition of charge between copper and oxygen but struggle to differentiate 
partition between planar and apical bonds~\cite{chen;prl91,chen;prl92,dalel;ijmpb04}. 

In Figure~\ref{fig;fig3}(a) we show $\eta_p$ on an expanded scale. To obtain an analytic 
form for the $p(x)$ calibration curve we have fit curves to the data in Fig.~\ref{fig;fig3}(a). 
Linear fit proved adequate resulting in $\eta_p=-0.45(7)x+1.00(1)$. This results in the 
calibration 
\begin{equation}
p(x) = x (1.00(1) - 0.45(7) x).
\label{eq;calib}
\end{equation}
The resulting fit to the data is shown in Fig.~\ref{fig;fig3}(a) and (b) and in 
the inset as a dashed line. Thus, at a strontium content of $x=0.125$ the 
doping in the planar orbitals is actually $p=0.119$, and 
conversely, the rational 1/8 filling occurs at $x=0.131$. Interestingly, the 
plateau in resistivity that was observed by Komiya~\etal\cite{komiy;prl05} 
and correlated with special behavior at 1/8 doping occurred at $x=0.13$, which 
would correspond to $p=0.125$ when corrected using our calibration curve. 
Note that, at all dopings, more than 85\% of doped charge goes into the 
planar orbitals, though it appears increasingly  in the apical bonds 
at higher doping. 

\section{
Conclusions
}

In summary, we have tested the hypothesis that doping in \lsco\ is occurring 
predominantly in \aporb\ orbitals~\cite{perry;prb02}, rather than the 
canonical picture of doping into the planar \plorb\ bands. 
Whilst the average structure evolves {\it qualitatively} as would be 
predicted according to \aporb\ doping picture, more quantitative analysis 
suggests that the canonical picture of one hole doped into the planar bands 
per strontium is actually rather close to the true situation.  
We present a correction factor quantifying the partition of doped charge 
between planar and apical bonds. 

\begin{acknowledgements} 
We would like to acknowledge M. J. Gutmann, A. Bianconi, N. L. Saini 
and P. G. Radaelli for help and discussions. We acknowledge beamtime on the 
instrument GEM at ISIS. This work was supported by NSF through grant 
DMR-0304391. SJB acknowledges support and hospitality from Universit{\`a} 
di Roma, ``La Sapienza", where part of the work was carried out.
\end{acknowledgements}


\begin{thebibliography}{10}

\bibitem{perry;prb02}
J.~K. Perry, J.~Tahir-Kheli, and W.~A. Goddard-III,
\newblock Phys. Rev. B {\bf 65}, 144501 (2002).


\bibitem{haske;prb97}
D.~Haskel, E.~A. Stern, D.~G. Hinks, A.~W. Mitchell, and J.~D. Jorgensen,
\newblock Phys. Rev. B {\bf 56}, R521 (1997).


\bibitem{radae;prb94i}
P.~G. Radaelli, D.~G. Hinks, A.~W. Mitchell, B.~A. Hunter, J.~L. Wagner, 
B.~Dabrowski, K.~G. Vandervoort, H.~K. Viswanathan, and J.~D. Jorgensen,
\newblock Phys. Rev. B {\bf 49}, 4163 (1994).


\bibitem{billi;cc04}
S.~J.~L. Billinge and M.~G. Kanatzidis
\newblock Chem. Commun. 749 (2004).


\bibitem{bozin;prb99}
E.~S. Bo{\v z}in, S.~J.~L. Billinge, G.~H. Kwei, and H.~Takagi,
\newblock Phys. Rev. B {\bf 59}, 4445 (1999).


\bibitem{bozin;phd03}
E.~S.~Bo{\v z}in,
\newblock Ph.D.\ thesis, Michigan State University, 2003.

\bibitem{larso;laur94}
A.C. Larson and R.B. Von Dreele, 
\newblock Los Alamos Nat. Lab. Rep. LAUR 86-748 (1994).

\bibitem{egami;b;utbp03}
T.~Egami and S.~J.~L. Billinge,
\newblock {\em Underneath the Bragg Peaks: Structural analysis of complex
  materials},
\newblock Pergamon Press, Elsevier, Oxford, England, 2003.

\bibitem{peter;jac00}
P.~F. Peterson, M.~Gutmann, {Th.~Proffen}, and S.~J.~L. Billinge,
\newblock J. Appl. Crystallogr. {\bf 33}, 1192 (2000).

\bibitem{proff;jac99}
{Th. Proffen} and S.~J.~L. Billinge,
\newblock J. Appl. Crystallogr. {\bf 32}, 572 (1999).

\bibitem{komiy;prl05}
S. Komiya, {H.-D.} Chen, {S.-C.} Zhang, and Y. Ando,
\newblock Phys. Rev. Lett. {\bf 94}, 207004 (2005).

\bibitem{hanag;nat04}
T. Hanaguri, C. Lupien, Y. Kohsaka, {D.-H.} Lee, M. Azuma, M. Takano, H. Takagi, and J. C. Davis, 
\newblock Nature {\bf 430}, 1001 (2004).

\bibitem{shann;ac69}
R.~D. Shannon and C.~T. Prewitt,
\newblock Acta Cryst. B {\bf 25}, 925 (1969).

\bibitem{brade;prb01}
M.~Braden, M.~Meven, W.~Reichardt, L.~Pintschovius, M.~T. Fernandez-Diaz, 
G.~Heger, F.~Nakamura, and T.~Fujita,
\newblock Phys. Rev. B {\bf 63}, 140510 (2001).

\bibitem{brown;acb85}
I.~D. Brown and D.~Altermatt,
%\newblock Acta Cryst. B {\bf 41}, 244 (1985).
\newblock Acta Cryst. B 41 (1985)  244.

\bibitem{brown;jssc89}
I.~D. Brown,
\newblock J. Solid State Chem. {\bf 82}, 122 (1989).

\bibitem{takhi;unpub}
J.~Tahir-Kheli,
\newblock (private communication).

\bibitem{chen;prl91}
C.~T. Chen, F. Sette, Y. Ma, M.~S. Hybertsen, E.~B. Stechel, W.~M.~C. Foulkes, 
M. Schluter, {S-W. Cheong}, A.~S. Cooper, {L.~W. Rupp, Jr.}, B. Batlogg, 
Y.~L. Soo, Z.~H. Ming, A. Krol, and Y.~H. Kao, 
\newblock Phys. Rev. Lett. {\bf 66}, 104 (1991).

\bibitem{chen;prl92}
C.~T. Chen, L.~H.~Tjeng, J. Kwo, H.~L. Kao, P. Rudolf, F. Sette, and R.~M. Fleming, 
\newblock Phys. Rev. Lett. {\bf 68}, 2543 (1992).

\bibitem{dalel;ijmpb04}
B. Dalela, S. Dalela, N.~L. Saini, R.~K. Singhal, C.~T. Chen, and K.~B. Garg, 
\newblock Int. J. Mod. Phys. B {\bf 18}, 2849 (2004).

\end{thebibliography}
\end{document}